\documentclass[twocolumn,aps,prb,showpacs,amsmath,superscriptaddress,longbibliography,notitlepage]{revtex4-2}
\usepackage{amssymb}
\usepackage{mathrsfs}
\usepackage{graphicx}
\usepackage{float}
\usepackage[caption=false]{subfig}
\usepackage[pdftex,colorlinks=red]{hyperref}
\usepackage{makecell}
\usepackage{epstopdf}
\usepackage[flushleft]{threeparttable}
\usepackage{booktabs}
\usepackage{amsmath}
\usepackage[normalem]{ulem}
\usepackage{verbatim}
\usepackage{xcolor}
\usepackage{bm}

\bibpunct{[}{]}{,}{n}{}{}

\begin{document}

\title{Hierarchical Topological Protection and Subsymmetry-Driven Edge-State Switching}
\author{Shu Long}
\affiliation{Quantum Science Center of Guangdong-Hong Kong-Macao Greater Bay Area (Guangdong), Shenzhen, China}
\affiliation{Guangdong Provincial Key Laboratory of Quantum Metrology and Sensing $\&$ School of Physics and Astronomy, Sun Yat-Sen University (Zhuhai Campus), Zhuhai 519082, China}
\author{Rijia Lin}
\affiliation{Guangdong Provincial Key Laboratory of Quantum Metrology and Sensing $\&$ School of Physics and Astronomy, Sun Yat-Sen University (Zhuhai Campus), Zhuhai 519082, China}
\author{Hui-Qiang Liang} \email{huiqiang.liang1091@gmail.com}
\affiliation{Quantum Science Center of Guangdong-Hong Kong-Macao Greater Bay Area (Guangdong), Shenzhen, China}
\affiliation{Department of Physics, Shandong University, Jinan 250100, China}
\author{Linhu Li}  \email{lilinhu@quantumsc.cn}
\affiliation{Quantum Science Center of Guangdong-Hong Kong-Macao Greater Bay Area (Guangdong), Shenzhen, China}

\begin{abstract}
	We reveal a hierarchy of topological protection in which multiple symmetries and their reduced forms provide distinct mechanisms for protecting topological edge states. 
	In Hermitian Hamiltonians obtained by doubling chiral non-Hermitian systems, a structural and an inherited chiral symmetry provide complementary channels of topological protection. 
	By selectively breaking either symmetry, we demonstrate their distinct roles in protecting edge zero modes, establishing different levels of protection under symmetry reduction. 
	Remarkably, even when both full chiral symmetries are broken, fixed subsymmetry constraints allow the protected zero-mode configurations to switch between opposite edges and change in number across the topological phase boundaries of the symmetry-unbroken system. 
	Our results establish a hierarchical framework for topological protection, revealing a route to controlling edge-state configurations via reduced symmetry constraints.
\end{abstract}

\date{\today}
\maketitle

\emph{Introduction---}
Symmetry plays a fundamental role in the study of topological phases of matter,
governing the classification of bulk topology and the protection of topological boundary states~\cite{RevModPhys.82.3045,RevModPhys.83.1057,RevModPhys.88.035005,Senthil2015,PhysRevLett.89.077002,PhysRevB.90.115207,PhysRevB.107.075422,PhysRevB.96.035149,PhysRevB.84.125132,PhysRevB.97.115143,PhysRevB.89.085111}. 
The recently introduced subsymmetry (SubSy) has further broadened this framework, 
revealing that topological boundary states can survive within selected subspaces even when the corresponding full symmetry is broken~\cite{wang2023sub,PhysRevB.110.035424,10.1021/acsphotonics.4c00600,cheng2026,xia2026control}. 
This concept has since been extended to topological superconductors~\cite{PhysRevResearch.6.033323} and higher-order topological systems~\cite{kang2026higher}, as well as nonzero-energy boundary states~\cite{li2026general}.
While these advances have revealed a richer landscape of topological protection, an important question arises when multiple protecting symmetries coexist: 
how do their distinct protection mechanisms evolve under symmetry reduction, and what boundary-state structures can emerge from the residual constraints?

Non-Hermitian topology provides a natural setting for exploring this question. 
Unlike Hermitian systems, non-Hermitian Hamiltonians support point-gap topology characterized by complex spectral winding, in addition to symmetry-constrained band topology~\cite{PhysRevX.8.031079,PhysRevX.9.041015,PhysRevLett.124.086801,Zhang2020correspondence,annurev:/cont,PhysRevLett.132.136401,PhysRevLett.121.086803,PhysRevLett.123.066404,lin2023topological,nnpf-8pxq,Liang_2022}. 
Through Hermitian doubling, a non-Hermitian Hamiltonian can be mapped onto a Hermitian one possessing a chiral symmetry from the doubling structure, which establishes a correspondence between non-Hermitian spectral winding and Hermitian topological invariants~\cite{PhysRevX.8.031079,PhysRevX.9.041015,PhysRevB.99.125103,PhysRevB.104.035424,PhysRevLett.123.206404}. 
Notably, the original non-Hermitian Hamiltonian can itself support extra symmetries that protect band topology therein, which may also be inherited by the doubled Hermitian system. The resulting structural and inherited chiral symmetries coexist and can give rise to distinct topological invariants. Although their coexistence allows a symmetry-resolved description of the topology, how the resulting topological protection evolves when either symmetry is independently broken, and whether their residual subsymmetries can provide more flexible boundary-state protection, remain to be explored.

\begin{figure}[t]
	\centering
	\includegraphics[width=1\linewidth]{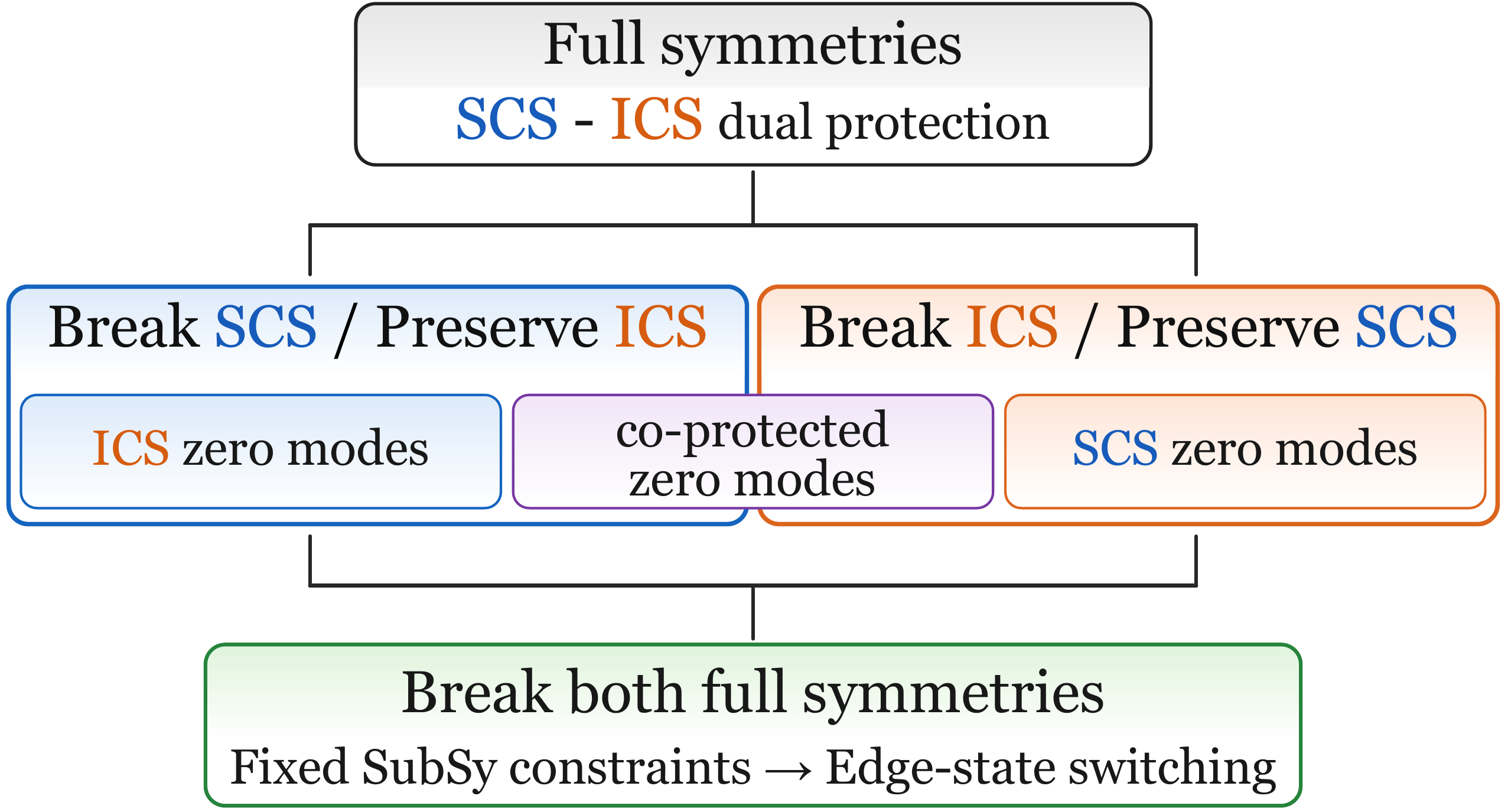}
	\caption{Hierarchical topological protection under symmetry reduction.
The coexistence of a structural and an inherited chiral symmetries (SCS and ICS) provides complementary channels of topological protection. Breaking either symmetry leaves zero edge modes protected by the remaining symmetry, including co-protected modes that survive the breaking of either symmetry. When both full chiral symmetries are broken, residual subsymmetry (SubSy) constraints can selectively preserve zero edge modes, enabling topology-controlled edge-state switching.}
	\label{fig:sketch}
\end{figure}

In this Letter, we uncover a hierarchical topological protection of edge states in Hermitian systems constructed by doubling chiral non-Hermitian Hamiltonians, as summarized in Fig.~\ref{fig:sketch}. 
We show that a structural and an inherited chiral symmetries (SCS and ICS) jointly organize the topology into distinct winding channels, while selectively breaking either symmetry removes the corresponding topological structure, without necessarily destroying the topology protected by the remaining chiral symmetry. Building on this structure, we further construct two distinct subsymmetries that remain intact even when both full chiral symmetries are broken. Remarkably, these fixed SubSy constraints selectively protect zero modes at opposite edges across  different topological regimes of the symmetry-unbroken system, enabling edge-state switching without modifying the protecting symmetry conditions. 
Our results establish a hierarchy of topological protection encompassing multiple full symmetries, individual symmetry protection, and residual SubSy protection. 
This hierarchy not only clarifies the distinct roles of multiple protecting symmetries but also provides a mechanism for controlling the edge localization of protected zero modes.

\emph{General structure of hierarchical topology under multiple symmetries---}
For an arbitrary non-Hermitian Hamiltonian $Q$, a doubled Hermitian Hamiltonian can be constructed as\cite{PhysRevX.8.031079,PhysRevX.9.041015,PhysRevB.99.125103,PhysRevB.104.035424,PhysRevLett.123.206404}
\begin{equation}\label{eq_QH}
      H = \begin{pmatrix}
             0 & Q  \\
            Q^{\dagger}  & 0
            \end{pmatrix},
    \end{equation}
which satisfies a SCS arising from the doubling construction,
\begin{equation}\label{eq_s1_sym}
      \mathcal{S}_{H} H \mathcal{S}_{H}^{-1} = -H.
\end{equation}
For the two-band model considered below, the SCS operator is explicitly given by $\mathcal S_H=\tau_z\sigma_0$, where $\sigma_0$ denotes the identity matrix in the original pseudospin space.
In one dimension (1D), $S_{H}$ protects $Z$-type topology 
of the doubled Hermitian Hamiltonian $H$, characterized by the non-Hermitian spectral winding of $Q$ \cite{PhysRevX.8.031079}. 
On the other hand, a multi-band $Q(k)$ itself may also support non-Hermitian band topology.
For example, a chiral symmetry of $Q(k)$ is naturally inherited by the doubled Hamiltonian $H$,
\begin{align}
\mathcal{S}^{\sigma}_{Q} Q(k) [\mathcal{S}^{\sigma}_{Q}]^{-1}=-Q(k)\Rightarrow \mathcal{S}_{Q} H(k) \mathcal{S}_{Q}^{-1}=-H(k)
\end{align}
with $\mathcal{S}_{Q}=\tau_0\mathcal{S}^{\sigma}_{Q}$,
which we dub as the ICS.
For concreteness, we consider a two-band chiral Hamiltonian of the form
\begin{align}
Q(k)	=
	\begin{pmatrix}
		0&q_+(k)\\
		q^*_-(k)&0
	\end{pmatrix},\quad q_\pm=q_{\rm H}\pm q_{\rm aH},\label{eq:Q_general}
\end{align}
where $q_{\rm H}$ and $q_{\rm aH}$ are the Hermitian and anti-Hermitian parts of $Q$, respectively,
and the symmetry operator is given by $S^{\sigma}_Q=\sigma_z$, the third Pauli matrix acting in the internal space of $Q$.
Accordingly, a topological invariant can be assigned  to each symmetry as
   \begin{equation}
   	\nu_{\alpha} = \frac{1}{2\pi i} \int_{-\pi}^{\pi} {\rm d}k \, \partial_k \ln \det D_{\alpha}(k), \quad \alpha\in\{Q,H\},\label{eq:nu_H_Q}
   \end{equation}
where $ D_{\alpha}$ is obtained by transforming the Hamiltonian into off-diagonal form according to the symmetry,
   \begin{align}
   	  &\widetilde{H}_{\alpha}(k) = U_{\alpha}^{\dagger} H(k) U_{\alpha} = 
   	  \begin{pmatrix}
   	  	0 & D_{\alpha}(k) \\
   	  	D_{\alpha}^{\dagger}(k) & 0
   	  \end{pmatrix},\nonumber\\ &U^\dagger_\alpha S_\alpha U_\alpha={\rm diag}(1,1,-1,-1).
   \end{align}

Essentially, the two topological invariants defined in Eq.~\eqref{eq:nu_H_Q} describe different topological structures of the non-Hermitian generator $Q$.
Specifically, the doubled structure of Eq.~\eqref{eq_QH} ensures $D_H=Q$, and $\nu_H$ describes the spectral winding of $Q$ around a reference energy at zero. In our particular setting with $S_Q$,  $\nu_H$ further describes the energy vorticity of $Q$ \cite{PhysRevLett.120.146402,PhysRevLett.118.040401}, reflecting the braiding between its non-Hermitian bands \cite{PhysRevLett.126.010401};
while $\nu_Q$ corresponds to the PBC band topology of $Q$ \cite{PhysRevA.97.052115}, whose OBC correspondence can be violated due to the non-Hermitian skin effect \cite{PhysRevLett.121.086803}.
In contrast, in the Hermitian Hamiltonian $H$, $\nu_H$ and $\nu_Q$ provide two complementary channels that protect zero edge modes, while selectively breaking the two symmetries (SCS and ICS) for the full system or its subspaces further leads to a hierarchy of topological protection, as we will elaborate below.

\begin{figure}[t]
	\centering
	\includegraphics[width=1\linewidth]{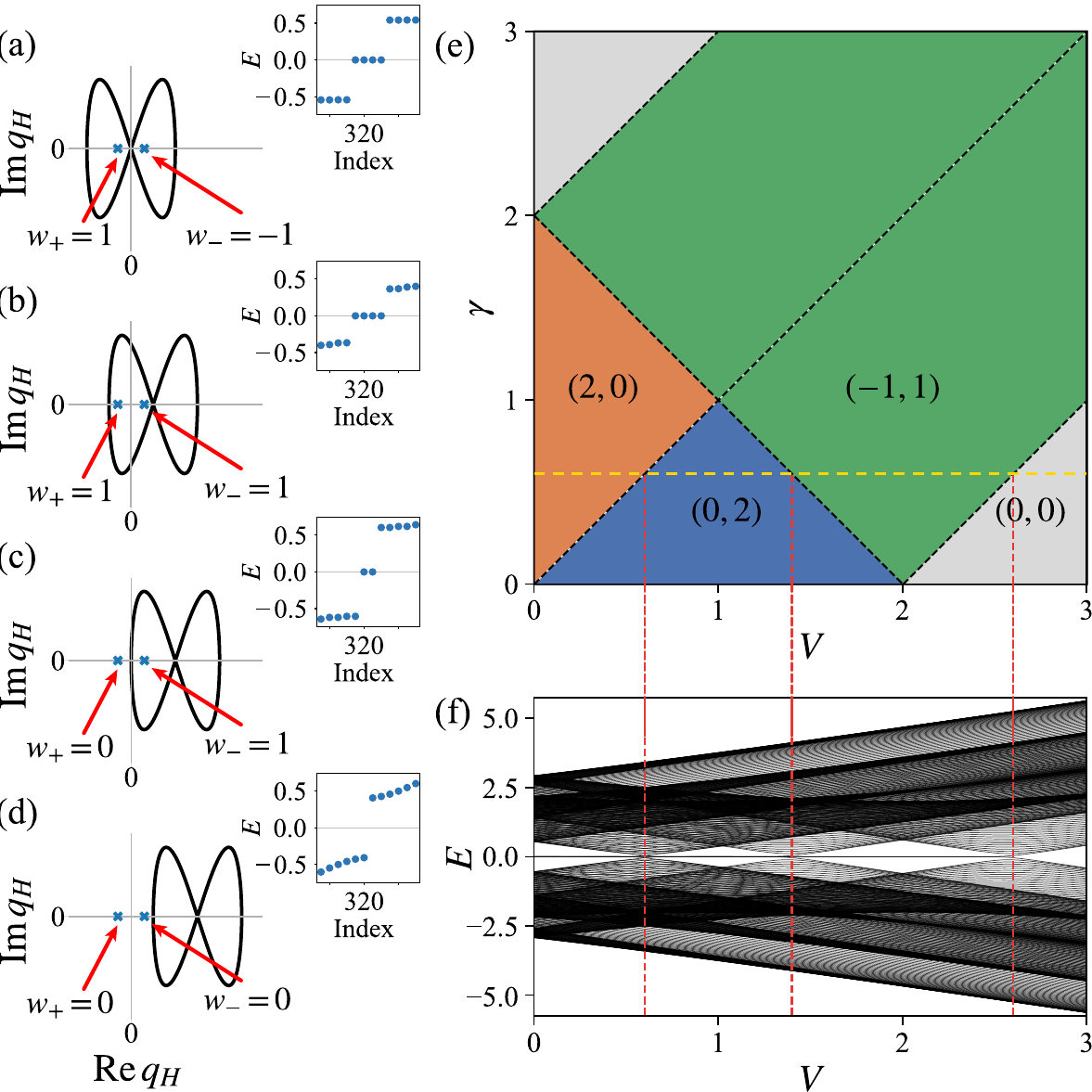}
	\caption{(a) to (d) Trajectories of $q_H$ in the complex plane at $\gamma=0.6$, for $V=0,1,2,3$, respectively. The blue crosses indicate the two EPs, whose positions are determined by $\gamma$. The insets show a subset of the sorted eigenenergies of the doubled Hermitian Hamiltonian. (e) Phase diagram of the system in the $(\gamma,V)$ parameter space, where different phases are labeled by $(\nu_H,\nu_Q)$. (f) Eigenenergies of the doubled Hermitian Hamiltonian as a function of $V$ at fixed $\gamma=0.6$. The system size is $N=160$.}
	\label{fig:fig1tempnew}
\end{figure}

\emph{Minimal model and hierarchical topological protection---}
As a demonstration, we consider a non-Hermitian generator given by Eq.~\eqref{eq:Q_general} with
\begin{align}
q_{\rm H}=z^2-z^{-2}-i(z-z^{-1})+V,\quad q_{\rm aH}=\gamma,\label{eq:q}
\end{align}
where $z=e^{ik}$.
The full Hamiltonian is given by
\begin{equation}
	H(k)=(V+2\sin k)\tau_x\sigma_x-2\sin2k\,\tau_x\sigma_y-\gamma\tau_y\sigma_y,
\end{equation}
constituting a minimal model that supports all the four possibilities regarding trivial and nontrivial topology of $\nu_H$ and $\nu_Q$. To see this,
we note that in the presence of both SCS and ICS, the two topological invariants can be further decomposed as (see Sec. I in Supplemental Material \cite{SuppMat})
\begin{align}
\nu_H=w_+ - w_-,\quad\nu_Q=w_++w_-,
\end{align}
with
\begin{equation}\label{eq:wpm}
	w_\pm=\frac{1}{2\pi i}\int_{-\pi}^{\pi}{\rm d}k\,\partial_k\ln q_\pm(k)
\end{equation}
describing the winding of $q_{\rm H}$ trajectory around the two exceptional points (EPs) of $Q$ given by $q_{\rm H}=\pm q_{\rm aH}$~\cite{PhysRevA.97.052115,PhysRevB.100.075403}. 
The explicit form of $q_{\rm H}$ in Eq.~\eqref{eq:q} gives a figure-eight trajectory when $k$ varies from $0$ to $2\pi$, 
which is the simplest 1D geometry supporting both positive and negative $w_\pm$ and hence different combinations of trivial and nontrivial $\nu_H$ and $\nu_Q$, as visualized in Figs.~\ref{fig:fig1tempnew}(a) to~\ref{fig:fig1tempnew}(d).

The phase boundaries can be obtained analytically from the condition that the trajectory of $q_{\rm H}(k)$ passes through either of the two EPs at $q_{\rm H}=\pm\gamma$, yielding 
\begin{equation}
	\gamma=V,\qquad \gamma=V+2,\qquad \gamma=|V-2|.
\end{equation}
These analytical boundaries exactly coincide with the phase boundaries obtained from the winding numbers $\nu_H$ and $\nu_Q$, as shown in Fig.~\ref{fig:fig1tempnew}(e).
In Fig.~\ref{fig:fig1tempnew}(f), we present the eigenenergy spectrum of the doubled Hermitian Hamiltonian $H$ as a function of $V$ at fixed $\gamma$, corresponding to a one-dimensional cut through the phase diagram in Fig.~\ref{fig:fig1tempnew}(e). 
The vertical dashed lines indicate the intersections with the analytical phase boundaries, across which the number of zero edge modes changes.
The distinct topological phases can be identified by their winding numbers and the corresponding numbers of edge zero modes.
Namely, the system supports four zero edge modes in the phases with $(\nu_H,\nu_Q)=(2,0)$ and $(0,2)$, 
indicating distinct topological protection mechanisms despite the same number of edge zero modes [also see insets of Figs.~\ref{fig:fig1tempnew}(a) and ~\ref{fig:fig1tempnew}(b)]. On the other hand, only two zero edge modes emerge in the phase with $(\nu_H,\nu_Q)=(-1,1)$ [also see inset of Fig.~\ref{fig:fig1tempnew}(c)], suggesting the co-protection of the same edge states by topology arising from both symmetries. 

\begin{figure}
	\centering
	\includegraphics[width=1\linewidth]{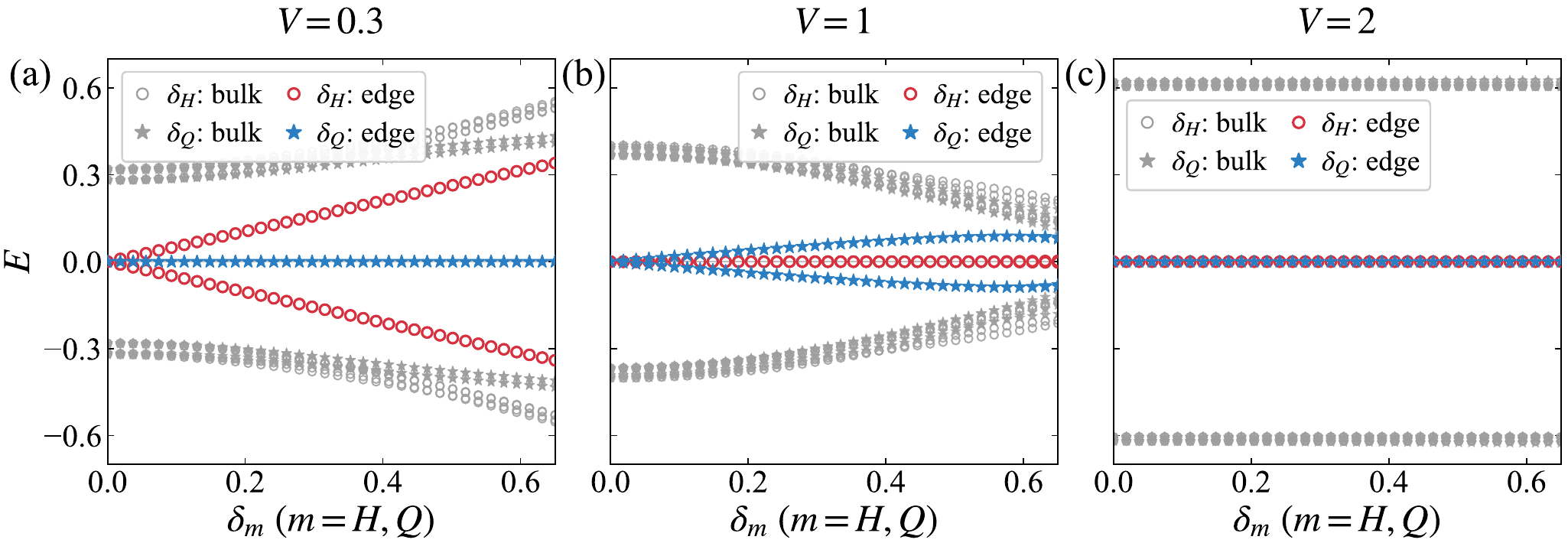}
	\caption{Evolution of the low-energy OBC spectra of the Hamiltonian $H+\Delta^{\rm full}_m$ ($m=H$ or $Q$) under selective symmetry breaking at (a) $V=0.3$, (b) $V=1$, and (c) $V=2$. 
		We introduce the two perturbations of Eq.~\eqref{eq:delta}, respectively, to break $\mathcal S_H$ and $\mathcal S_Q$ separately.
		Open circles (stars) represent the eigenvalues obtained by varying $\delta_H$ ($\delta_Q$). Red and blue markers indicate edge-localized states, while gray markers represent bulk states. For clarity, only the four bulk eigenvalues closest to zero on each side of the spectrum are displayed for each perturbation. The parameters are $\gamma=0.6$ and $N=160$.}
	\label{fig:fig2tempnew}
\end{figure}

These distinct topological origins of the two-fold and four-fold degenerate edge states can be further clarified by reducing the symmetries of the full Hamiltonian, constituting a hierarchical topological protection of the edge states. Explicitly, as the two chiral symmetries act on different subspaces, each of them can be removed by adding the third Pauli's matrix or the identity matrix solely in the corresponding subspace, without breaking the other.
As examples, we introduce single-symmetry breaking terms 
\begin{align}
\Delta^{\rm full}_H=\delta_H \cos k \tau_0\sigma_x,\quad \Delta^{\rm full}_Q=-\delta_Q\cos k\tau_y\sigma_z,\label{eq:delta}
\end{align}
which break each of $S_H$ and $S_Q$, respectively.
We choose hopping-type rather than on-site perturbations, as the latter 
can yield vanishing first-order energy corrections due to the microscopic structures of the Hamiltonian and edge states (see Sec. II in Supplemental Material \cite{SuppMat}).
In Fig. \ref{fig:fig2tempnew}, we present typical cases of the three topologically nontrivial regimes under different symmetry breaking.
It is seen in Fig. \ref{fig:fig2tempnew}(a) that the 4-fold degeneracy in the $(2,0)$ phase is lifted by $\delta_H$, but remains under $\delta_Q$.
For the $(0,2)$ phase, the two terms have precisely opposite effects on the edge states [Fig. \ref{fig:fig2tempnew}(b)]. 
Thus, each of these two phases is protected by a single chiral symmetry of the system.
In contrast, for the $(-1,1)$ phase shown in Fig. \ref{fig:fig2tempnew}(c), the edge states remain degenerate at zero energy with either symmetry being broken, demonstrating that the edge zero modes remain protected upon breaking either chiral symmetry.

\emph{SubSy-driven edge-state switching---}
The hierarchical topological protection enabled by multiple symmetries facilitates schemes for engineering and manipulating topological edge states in distinct subspaces. Fig.~\ref{fig:fig2tempnew} already provides an example where the symmetry is fully broken in one of the two pseudospin subspaces.
Here we further adopt the SubSy formalism \cite{wang2023sub,PhysRevB.110.035424,PhysRevResearch.6.033323} and break both chiral symmetries in different internal subspaces of the system~\cite{SuppMat}. Explicitly, we introduce two SubSy perturbations,
\begin{align}
\Delta^{\rm sub}_H={\eta_H\cos k}\tau_+\sigma_x,\quad 
\Delta^{\rm sub}_Q={\eta_Q\cos k}\tau_x\sigma_-,\label{eq:sub_breaking}
\end{align}
where $\tau_\pm=(\tau_0\pm\tau_z)/2$ and $\sigma _\pm=(\sigma_0\pm\sigma_z)/2$.
Thus, the perturbed Hamiltonian 
\begin{align}
H^{\rm sub}=H+\Delta^{\rm sub}_H+\Delta^{\rm sub}_Q
\end{align}
satisfies two subsymmetries,
\begin{align}
	(\mathcal S_H H^{\rm sub}\mathcal S_H^{-1}+H^{\rm sub})P_{H,-}=0, \label{eq:subsys1}\\
	(\mathcal S_Q H^{\rm sub}\mathcal S_Q^{-1}+H^{\rm sub})P_{Q,+}=0, \label{eq:subsys2}
\end{align}
where the projection operators are defined as
\begin{equation}
    P_{\alpha,\pm}=\frac{I\pm\mathcal S_\alpha}{2}
\end{equation}
with $\alpha=H$ or $Q$.
Note that the edge-state switching introduced below is insensitive to the explicit forms of Eq.~\eqref{eq:sub_breaking}, provided the SubSy constraints are satisfied. Another example with edge-state switching driven by disordered SubSy perturbations is shown in Supplemental Materials~\cite{SuppMat}.

\begin{figure}
	\centering
	\includegraphics[width=1\linewidth]{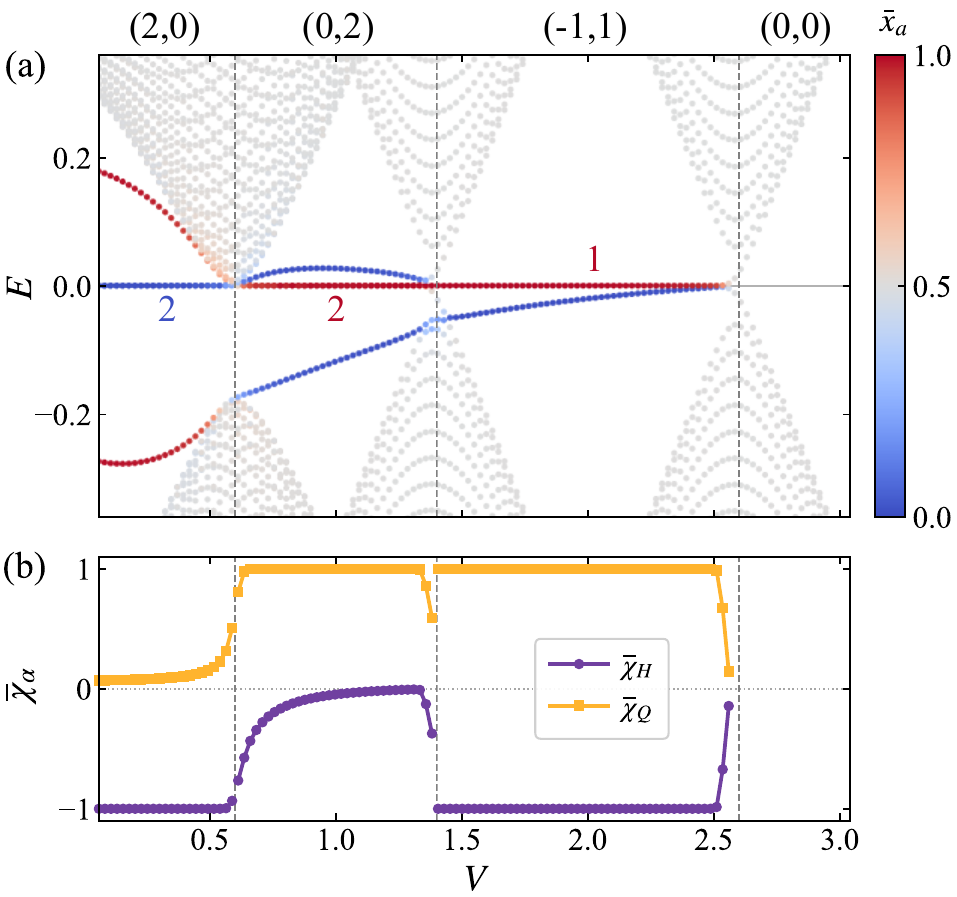}
	\caption{(a) Low-energy OBC spectrum of the Hamiltonian with SubSy perturbations, $H^{\rm sub}$, with $\gamma=0.6$, $\eta_H=0.4$, and $\eta_Q=0.3$. The color represents the normalized mean position $\bar x_a$ of each eigenstate, ranging from 0 (left edge) to 1 (right edge). 
    The topological invariants $(\nu_H,\nu_Q)$ of the original doubled Hermitian Hamiltonian $H$ are indicated on top of the figure.
    The number of zero edge modes of $H^{\rm sub}$ in different topological regimes is also indicated in the figure.
    A switching of the zero edge modes between the two edges is observed at the transition between regimes with $(\nu_H,\nu_Q)=(2,0)$ and $(0,2)$.
	The vertical dashed lines mark the zero-energy bulk gap-closing points at $V=0.6,1.4$, and $2.6$. 
	(b) Average chiralities of the edge states, as defined in Eq.~\eqref{eq:ave_chi}, identifying the protection by different symmetries of the zero edge modes.
	The system contains $L=300$ unit cells.}
	\label{fig:fig4temp}
\end{figure}

An important consequence of Eqs.~\eqref{eq:subsys1} and~\eqref{eq:subsys2} is that the two SubSy conditions impose different restrictions on the internal structure of zero edge modes. 
The $P_{H,-}$-SubSy preserves the zero diagonal block associated with the $\tau_-$ sector, whereas the $P_{Q,+}$-SubSy preserves the corresponding zero block within the $\sigma_+$ pseudospin sector. 
Since the original Hamiltonian supports edge states with different spatial localizations and internal structures in different topological phases, the same projected symmetry constraints need not protect zero modes at the same edge in every phase. 
Instead, they can selectively preserve different subsets of the original zero edge modes as the system parameters are varied.

To demonstrate this effect, we calculate the OBC energy spectrum of the perturbed Hamiltonian $H^{\rm sub}$ in Fig.~\ref{fig:fig4temp}(a). 
The color of each eigenvalue represents the normalized mean position of its corresponding eigenstate, given by
\begin{equation}
	\bar{x}_a=
	\frac{\displaystyle\sum_{j=1}^{N}\frac{j-1}{N-1}\sum_{\mu=1}^{4}|\psi_{a,j\mu}|^2}
	{\displaystyle\sum_{j=1}^{N}\sum_{\mu=1}^{4}|\psi_{a,j\mu}|^2},
\end{equation}
where $\psi_{a,j\mu}$ denotes the wave-function amplitude of the $a$-th eigenstate at unit cell $j$ and internal degree of freedom $\mu$, and  $N$ is the total number of unit cells. 
By construction, $\bar{x}_a\in[0,1]$, with values close to 0 and 1 indicating that the eigenstates are localized at the left and right edges, respectively.
It is seen that as $V$ varies, the system supports three nontrivial regimes exhibiting distinct zero-mode configurations, and a trivial regime without any edge states.
The boundaries between these regimes coincide with the topological phase boundaries of the unperturbed Hamiltonian $H$, determined by
\begin{align}
{\rm det} [H]={\rm det} [H^{\rm sub}]=|q_+q_-|^2.
\end{align}
Physically, this is because the zero edge modes origin from the topological edge states of $H$ in the subspaces where either SCS or ISC holds,
allowing the edge-state evolution to be traced across the original topological phase boundaries characterized by $(\nu_H,\nu_Q)$.

Remarkably, in the two topological regimes with $(\nu_H,\nu_Q)=(2,0)$ and $(0,2)$ of the unperturbed Hamiltonian $H$, only two of the four edge states remain at zero energy, whose spatial distributions switch between the two edges across the transition [Fig.~\ref{fig:fig4temp}(a)].
This is because in these two regimes, the zero-energy condition is protected by distinct chiral symmetries $S_H$ and $S_Q$,
which survive in different pseudospin sectors $\tau_-$ and $\sigma_+$, respectively.
To see this,
in Fig.~\ref{fig:fig4temp}(b), we demonstrate their average chiralities regarding SCS and ICS, defined as 
\begin{align}
	\overline{\chi}_\alpha=\frac{1}{N_0}\sum_{a=1}^{N_0}\langle\psi_a|\mathcal S_\alpha|\psi_a\rangle,\label{eq:ave_chi}
\end{align}
with $|\psi_a\rangle$ denoting normalized zero-energy eigenstates and $N_0$ their number. In the $(2,0)$ regime, $\overline{\chi}_H\approx-1$ indicating that the zero-energy states are also eigenstates of $S_H$, carrying negative $\tau$-chirality; while $\overline{\chi}_Q\approx0$ indicating a vanishing $\sigma$-chirality of the zero-mode subspace.
In contrast, the $(0,2)$ regime supports zero-energy states with positive $\sigma$-chirality and zero net $\tau$-chirality.
In other words, the switching of edge states represents a SubSy-driven topological phenomenon, facilitated by the SCS and ICS in different subspaces of the complete system.

Finally, the third topological regime with $(\nu_H,\nu_Q)=(-1,1)$ is seen to support a single zero edge mode localized at the right edge ($\bar{x}_a\approx1$), as shown in Fig.~\ref{fig:fig4temp}(a).
To understand this, we note that since $\mathcal S_H$ and $\mathcal S_Q$ commute, their projection operators also commute, allowing us to define the common projection
\begin{equation}
	P_{H,-}P_{Q,+}=\operatorname{diag}(0,0,1,0).
\end{equation}
This projector selects the $\tau_-\sigma_+$ subspace protected by both subsymmetries.
In the unperturbed Hamiltonian $H$, the edge states in this regime are co-protected by both SCS and ICS~[see Fig.~\ref{fig:fig2tempnew}(c)], thus possessing unitary chirality regarding each of them. In particular, the zero mode residing in the $\tau_-\sigma_+$ subspace remains protected against SubSy perturbations.
Indeed, as demonstrated in Fig.~\ref{fig:fig4temp}(b), the zero-energy state in this regime possesses $\overline{\chi}_H=-1$ and $\overline{\chi}_Q=1$, consistent with its ummunity to both SubSy perturbations in Eq.~\eqref{eq:sub_breaking}.

\emph{Summary---}
In this work, we have demonstrated a hierarchy of topological protection arising from multiple symmetries and their reduced forms in Hermitian Hamiltonians obtained by doubling chiral non-Hermitian systems. The structural and inherited chiral symmetries provide complementary channels of topological protection, whose distinct roles are revealed by selectively breaking either symmetry while preserving the other. Depending on the bulk topology, edge zero modes can be protected by either chiral symmetry or co-protected by both. We further extend this hierarchy to SubSy protection, where both full chiral symmetries are broken while residual SubSy constraints selectively preserve edge zero modes. Under these fixed constraints, the protected edge-state configurations switch between opposite edges and change in number across different topological regimes. Our results reveal how multiple symmetries and their reduced forms organize topological protection and facilitate topology-controlled manipulation of boundary states. Such topology-controlled switching could be probed in engineered photonic or circuit lattices by tuning bulk couplings while monitoring the spatial localization of boundary modes.

\emph{Acknowledgement---}
This work was supported by the National Natural Science Foundation of China (Grant No. 12474159) and the Guangdong Provincial Quantum Science Strategic Initiative (Grants No. GDZX2504003 and No. GDZX2504006).

\clearpage

\begin{widetext}
	\begin{center}
		\textbf{\large Supplemental Materials}
	\end{center}
	
	\tableofcontents
	\setcounter{secnumdepth}{2}
	\setcounter{equation}{0} \setcounter{figure}{0} 
	\setcounter{table}{0} %
	\renewcommand{\theequation}{S\arabic{equation}} 
	\renewcommand{\thefigure}{S%
		\arabic{figure}} \renewcommand{\bibnumfmt}[1]{[S#1]} 
	
\section{Winding numbers of the off-diagonal components}\label{appa}

The non-Hermitian Hamiltonian in Eq.~\eqref{eq:q} can be written as
\begin{equation}
	Q(k)=
	\begin{pmatrix}
		0&q_+(k)\\
		q_-^*(k)&0
	\end{pmatrix},
\end{equation}
where
\begin{equation}\label{eq:qpm}
	q_\pm(k)=V+2\sin k+2i\sin 2k\pm\gamma.
\end{equation}
The corresponding doubled Hermitian Hamiltonian in Eq.~\eqref{eq_QH} reads
\begin{equation}
	H(k)=
	\begin{pmatrix}
		0&0&0&q_+(k)\\
		0&0&q_-^*(k)&0\\
		0&q_-(k)&0&0\\
		q_+^*(k)&0&0&0
	\end{pmatrix}.
\end{equation}

Since $q_\pm(k)$ are complex functions, their trajectories form closed loops in the complex plane as $k$ varies over the Brillouin zone. We define the corresponding winding numbers as
\begin{equation}\label{eq:wpm}
	w_\pm=\frac{1}{2\pi i}\int_{-\pi}^{\pi}{\rm d}k\,\partial_k\ln q_\pm(k).
\end{equation}

For the first chiral symmetry
\begin{equation}
	\mathcal S_H=\tau_z\sigma_0={\rm diag}(1,1,-1,-1).
\end{equation}
the original basis is already the corresponding chiral basis, and the off-diagonal block is simply $D_H(k)=Q(k)$. Therefore,
\begin{equation}
	\begin{aligned}
		\nu_H
		&=\frac{1}{2\pi i}\int dk\,\partial_k\ln\det D_H(k)
		\\
		&=\frac{1}{2\pi i}\int dk\,\partial_k\ln[-q_+(k)q_-^*(k)]
		\\
		&=w_+ + w[q_-^*]
		\\
		&=w_+-w_-.
	\end{aligned}
	\label{eq:nuH1_wpm}
\end{equation}

For the second chiral symmetry
\begin{equation}
	\mathcal S_Q=\tau_0\sigma_z={\rm diag}(1,-1,1,-1),
\end{equation}
we rearrange the basis by the unitary permutation matrix
\begin{equation}\label{eq:U2}
	U_Q=
	\begin{pmatrix}
		1&0&0&0\\
		0&0&1&0\\
		0&1&0&0\\
		0&0&0&1
	\end{pmatrix},
\end{equation}
such that
\begin{equation}
	U_Q^\dagger\mathcal S_QU_Q=
	\begin{pmatrix}
		I_2&0\\
		0&-I_2
	\end{pmatrix}.
\end{equation}
In this basis,
\begin{equation}
	U_Q^\dagger H(k)U_Q=
	\begin{pmatrix}
		0&D_Q(k)\\
		D_Q^\dagger(k)&0
	\end{pmatrix},
\end{equation}
with
\begin{equation}\label{eq:D2_original}
	D_Q(k)=
	\begin{pmatrix}
		0&q_+(k)\\
		q_-(k)&0
	\end{pmatrix}.
\end{equation}
It follows that
\begin{equation}
	\begin{aligned}
		\nu_Q
		&=\frac{1}{2\pi i}\int dk\,\partial_k\ln\det D_Q(k)
		\\
		&=\frac{1}{2\pi i}\int dk\,\partial_k\ln[-q_+(k)q_-(k)]
		\\
		&=w_++w_-.
	\end{aligned}
	\label{eq:nuH2_wpm}
\end{equation}

	We consider four types of $\mathcal S_{H}$-breaking perturbations 
	\begin{equation}
		V=\delta_{0x}\tau_0\sigma_x+\delta_{0y}\tau_0\sigma_y+\delta_{zx}\tau_z\sigma_x+\delta_{zy}\tau_z\sigma_y.
	\end{equation}
	Since all four perturbations possess a diagonal block structure in the original doubling space, they can be written as
	\begin{equation}
		H'(k)=\begin{pmatrix}
			A_+&Q(k)\\
			Q^\dagger(k)&A_-
		\end{pmatrix},
	\end{equation}
	where $A_\pm=(\delta_{0x}\pm\delta_{zx})\sigma_x+(\delta_{0y}\pm\delta_{zy})\sigma_y.$
	Transform $H'(k)$ into the chiral basis of $\mathcal S_{Q}$, we have
	\begin{equation}
		U_Q^\dagger H'(k)U_Q=
		\begin{pmatrix}
			0&D'_Q(k)\\
			D_Q'^\dagger(k)&0
		\end{pmatrix},
		D'_Q(k)=
		\begin{pmatrix}
			a_+&q_+(k)\\
			q_-(k)&a_-
		\end{pmatrix}.
	\end{equation}
	where $a_\pm=\delta_{0x}\pm\delta_{zx}-i(\delta_{0y}\pm\delta_{zy})$.
	We calculate the determinant
	\begin{equation}
		\nu_Q'=\frac{1}{2\pi i}\int dk\,\partial_k\ln\left[a_+a_- -q_+(k)q_-(k)\right],
	\end{equation}
	and find that it contains additional correction terms compared with the determinant of the matrix in Eq.~\eqref{eq:D2_original}, so it cannot be directly mapped onto the $w_++w_-$ result in Eq.~\eqref{eq:nuH2_wpm}.

\section{Vanishing First-Order Corrections from On-Site Symmetry-Breaking Perturbations}
In this section we show that the edge states can remain at zero energy under certain on-site symmetry-breaking perturbations.
This is because the topological edge states are pseudospin polarized [Eqs.~\eqref{eq:zero_1} and~\eqref{eq:zero_2}]; and on-site perturbations only exchange different pseudospin components without disrupting the microscopic structure of the wave functions in the pseudospin subspaces [Eq.~\eqref{eq:perturbed_state}]. The symmetric structure of the unperturbed Hamiltonian further ensures shared eigen solutions in different pseudospin subspace [Eq.~\eqref{eq:shared_eigen}]. The combined effect of these properties causes the first-order correction term to vanish coincidentally.

To see this, we first consider the original non-Hermitian Hamiltonian
\begin{equation}
	Q(k)=
	\begin{pmatrix}
		0&q_+(k)\\
		q_-^*(k)&0
	\end{pmatrix},
\end{equation}
where $q_\pm(k)=q_{\rm H}(k)\pm\gamma$.
Through Hermitian doubling, one can construct
\begin{equation}
	H(k)=
	\begin{pmatrix}
		0&Q(k)\\
		Q^\dagger(k)&0
	\end{pmatrix}.
\end{equation}
To analyze the zero edge modes under open boundary conditions, we first write $Q$ in real space. 
By arranging all degrees of freedom of the $\sigma_+$ and $\sigma_-$ sublattices separately, we obtain
\begin{equation}
	Q_{\mathrm{OBC}}=
	\begin{pmatrix}
		0&F\\
		G&0
	\end{pmatrix},
\end{equation}
where $F$ and $G$ are the real-space hopping matrices corresponding to $q_+(k)$ and $q_-^*(k)$, respectively.
For a finite chain containing $L$ unit cells, they are both $L\times L$ matrices and can be expressed as
\begin{equation}
	F=\widehat q_{\rm H}+\gamma I_L,
	\qquad
	G=\widehat q_{\rm H}^{\dagger}-\gamma I_L.\label{Seq:FG}
\end{equation}
Here, $\widehat q_{\rm H}$ denotes the real-space hopping matrix obtained by inverse Fourier transforming $q_{\rm H}(k)$.
Since $\gamma$ is a real constant, the above two matrices satisfy
\begin{equation}
	G=F^\dagger-2\gamma I_L.
	\label{eq:B5}
\end{equation}

In the internal basis $(\tau_+\sigma_+,\tau_+\sigma_-,\tau_-\sigma_+,\tau_-\sigma_-)$, the full doubled Hermitian Hamiltonian can be written as
\begin{equation}
	H_{\mathrm{OBC}}=
	\begin{pmatrix}
		0&0&0&F\\
		0&0&G&0\\
		0&G^\dagger&0&0\\
		F^\dagger&0&0&0
	\end{pmatrix},
\end{equation}
To solve for the zero edge modes, we consider a general eigenstate of the form
\begin{equation}
	|\psi\rangle=
	\begin{pmatrix}
		u_A&u_B&v_A&v_B
	\end{pmatrix}^T,
	\label{eq:B7}
\end{equation}
with $(u,v)$ denoting the pseudospin components associated with $\tau_\pm$, and $(A,B)$ denoting the pseudospin components associated with $\sigma_\pm$.
Substituting it into the zero-energy eigen equation, the equation can be decomposed into four mutually independent conditions:
\begin{equation}
	\begin{aligned}
		Fv_B&=0,\qquad  Gv_A=0,\\
		G^\dagger u_B&=0,\qquad F^\dagger u_A=0.
	\end{aligned}
\end{equation}
This indicates that the zero-energy states of the full Hamiltonian can be constructed separately from the null-space solutions of the different coupling blocks.
On a semi-infinite chain, we consider respectively the states satisfying
\begin{equation}
	Fv_F=0,\qquad  Gv_G=0,
\end{equation}
and the states satisfying
\begin{equation}
	F^\dagger u_F=0,\qquad  G^\dagger u_G=0.
\end{equation}
We can choose $v_A=v_G,~u_A=u_B=v_B=0$ in Eq.~\eqref{eq:B7}, the first left-edge zero mode is constructed:
\begin{equation}
	|\psi_{L,1}\rangle=
	\begin{pmatrix}
		0&0&v_G&0
	\end{pmatrix}^T\label{eq:zero_1}
\end{equation}
Similarly, a second left-edge zero mode can also be constructed as
\begin{equation}
	|\psi_{L,2}\rangle=
	\begin{pmatrix}
		0&0&0&v_F
	\end{pmatrix}^T.\label{eq:zero_2}
\end{equation}

After obtaining these two zero-energy states, we further consider an on-site symmetry-breaking perturbation proportional to $\tau_0\sigma_x$:
\begin{equation}
	H_\delta^{\mathrm{on}}=\delta\tau_0\sigma_x.
\end{equation}
Since this perturbation does not involve hopping between different unit cells, its full real-space form is
\begin{equation}
	H_{\delta,\mathrm{OBC}}^{\mathrm{on}}=
	\delta
	\begin{pmatrix}
		0&I_L&0&0\\
		I_L&0&0&0\\
		0&0&0&I_L\\
		0&0&I_L&0
	\end{pmatrix}.
\end{equation}
According to degenerate perturbation theory, the first-order energy correction to the zero-energy states is determined by the perturbation matrix projected onto the zero-mode subspace:
\begin{equation}
	M_{ij}=\langle\psi_{L,i}|H_\delta^{\mathrm{on}}|\psi_{L,j}\rangle.
\end{equation}
Since $\sigma_x$ exchanges the $A$ and $B$ sublattices, while $|\psi_{L,1}\rangle$ occupies only the $A$ ($\sigma_+$) sublattice,
\begin{equation}
	\langle\psi_{L,1}|H_\delta^{\mathrm{on}}|\psi_{L,1}\rangle=0.
\end{equation}
Similarly,
\begin{equation}
	\langle\psi_{L,2}|H_\delta^{\mathrm{on}}|\psi_{L,2}\rangle=0.
\end{equation}
Therefore, whether first-order energy splitting occurs depends crucially on the off-diagonal matrix element between the two zero modes. To explicitly calculate this matrix element, we first act the perturbation on the second zero mode:
\begin{equation}
	H_\delta^{\mathrm{on}}|\psi_{L,2}\rangle=\delta
	\begin{pmatrix}
		0&I_L&0&0\\
		I_L&0&0&0\\
		0&0&0&I_L\\
		0&0&I_L&0
	\end{pmatrix}
	\begin{pmatrix}
		0\\0\\0\\v_F
	\end{pmatrix}\\[4pt]
	=
	\delta
	\begin{pmatrix}
		0\\0\\v_F\\0
	\end{pmatrix}.\label{eq:perturbed_state}
\end{equation}
Thus,
\begin{equation}
	\begin{aligned}
		\langle\psi_{L,1}|H_\delta^{\mathrm{on}}|\psi_{L,2}\rangle
		&=
		\begin{pmatrix}
			0&0&v_G^\dagger&0
		\end{pmatrix}
		\delta
		\begin{pmatrix}
			0\\0\\v_F\\0
		\end{pmatrix}\\[4pt]
		&=\delta\,v_G^\dagger v_F=\delta\langle v_G|v_F\rangle
	\end{aligned}
	\label{eq:B19}
\end{equation}
Using the relation between the two coupling blocks in Eq.~\eqref{eq:B5}, $Gv_G=0$ can be rewritten as
\begin{equation}
	(F^\dagger-2\gamma I_L)v_G=0,
\end{equation}
i.e.,
\begin{equation}
	F^\dagger v_G=2\gamma v_G.\label{eq:shared_eigen}
\end{equation}
On the other hand, since $Fv_F=0$, we have
\begin{equation}
	0=\langle v_G|Fv_F\rangle
	=\langle F^\dagger v_G|v_F\rangle
	=2\gamma\langle v_G|v_F\rangle.
\end{equation}
Therefore, when $\gamma\neq0$, 
\begin{equation}
	\langle v_G|v_F\rangle=0.
	\label{eq:B23}
\end{equation}
Substituting Eq.~\eqref{eq:B23} back into Eq.~\eqref{eq:B19}, we finally obtain
\begin{equation}
	\langle\psi_{L,1}|H_\delta^{\mathrm{on}}|\psi_{L,2}\rangle
	=\delta\langle v_G|v_F\rangle=0.
\end{equation}
Since the perturbation is Hermitian, the reverse matrix element also vanishes:
\begin{equation}
	\langle\psi_{L,2}|H_\delta^{\mathrm{on}}|\psi_{L,1}\rangle=0.
\end{equation}
Therefore, all matrix elements of the perturbation matrix are zero, and the first-order energy corrections to both zero-energy states vanish. Similarly, for the two zero modes located at the right edge, the energy corrections can also be shown to vanish.

\begin{figure}
	\centering
	\includegraphics[width=0.5\linewidth]{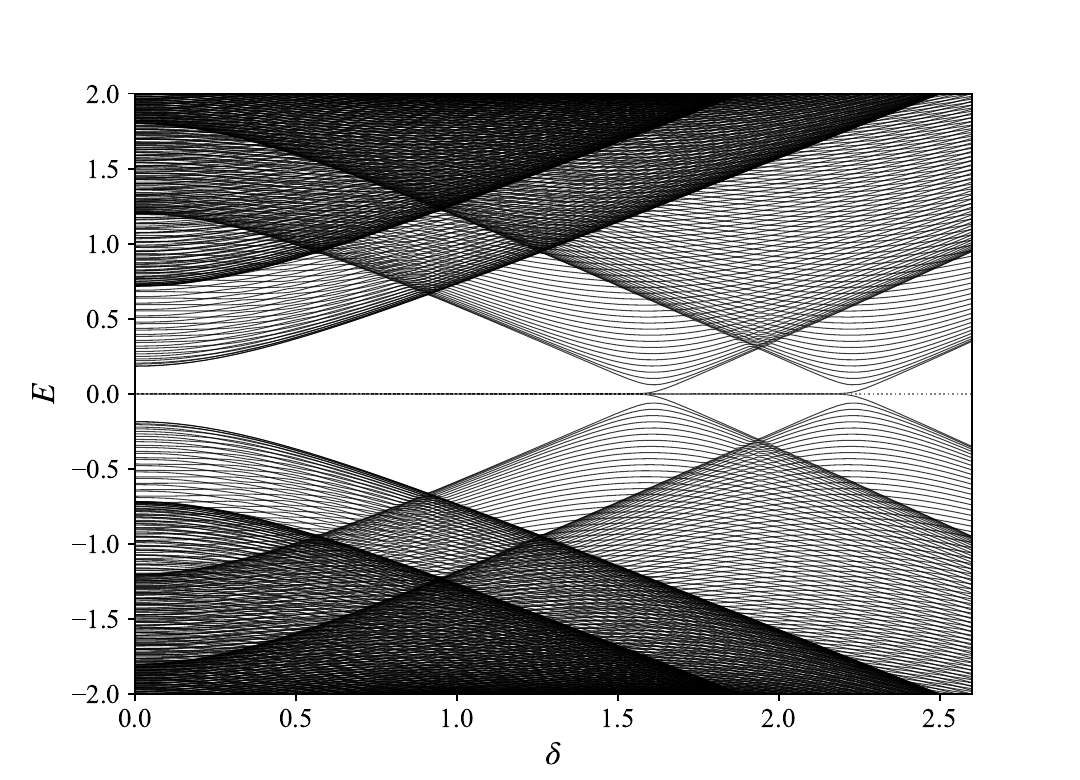}
	\caption{Low-energy OBC spectrum of the doubled Hermitian Hamiltonian as a function of the on-site symmetry-breaking strength $\delta$, with $V=0.3$, $\gamma=0.5$, and $L=300$. }
	\label{fig:figures1}
\end{figure}

To further verify this result beyond first-order perturbation theory, we numerically calculate the OBC spectrum under finite on-site perturbations. In the $(2,0)$ phase, the zero edge modes remain at zero energy despite the breaking of SCS, until the bulk gap closes at the topological phase transition [see Fig.~\ref{fig:figures1}].

\section{Construction of Two Subsymmetries}
In general, the topological protection of zero edge modes in chiral-symmetric Hermitian systems relies on the corresponding chiral symmetry. 
Once this symmetry is broken, the associated topological invariant may no longer be well defined, and the edge states are no longer guaranteed to remain at zero energy. 
However, a full chiral symmetry is not always necessary to protect a subset of zero edge modes. 
The recently proposed subsymmetry (SubSy) provides a framework in which certain edge states can remain protected even after the full symmetry is broken, provided that the Hamiltonian satisfies a residual symmetry constraint within a particular subspace~\cite{wang2023sub}.

For a chiral symmetry operator $\mathcal S_\alpha$, satisfying $\mathcal S_\alpha^2=I$, we define the projection operators
\begin{equation}
	P_{\alpha,\pm}=\frac{I\pm\mathcal S_\alpha}{2},
\end{equation}
which project onto the eigenspaces of $\mathcal S_\alpha$ with eigenvalues $+1$ and $-1$, respectively.
SubSy relaxes the symmetry requirement. 
For an operator $P_{\alpha,j}$ projected onto a specified chiral subspace, it is defined as
\begin{equation}
	\mathcal S_\alpha H\mathcal S_\alpha^{-1}P_{\alpha,j}=-HP_{\alpha,j},
\end{equation}
where $j=\pm$ specifies the projected chiral subspace.
Thus, unlike a full chiral symmetry, which forbids couplings within both chiral subspaces, SubSy only requires the absence of couplings within one selected subspace. 
Consequently, symmetry-breaking perturbations can be introduced into the complementary subspace without violating the projected symmetry condition. 
Such a constraint can preserve zero edge modes whose wave functions belong to the corresponding protected subspace, even when the full chiral symmetry is absent.

For the doubled Hermitian Hamiltonian considered in this work, the two original chiral symmetries are described by
\begin{equation}
	\mathcal S_H=\tau_z\sigma_0,\qquad
	\mathcal S_Q=\tau_0\sigma_z,
\end{equation}
where $\tau$ and $\sigma$ act on the doubling and pseudospin degrees of freedom, respectively. 
Specifically, we choose the projection operators
\begin{equation}
	P_{H,-}=\frac{I-\mathcal S_H}{2}
	=\frac{\tau_0-\tau_z}{2}\sigma_0,
\end{equation}
and
\begin{equation}
	P_{Q,+}=\frac{I+\mathcal S_Q}{2}
	=\tau_0\frac{\sigma_0+\sigma_z}{2}.
\end{equation}

To construct the first SubSy, we introduce a hopping-type perturbation $\Delta_H^{\rm sub}(k)=\eta_H\cos k\,\tau_+\sigma_x$, which acts exclusively within the $\tau_+$ doubling sector. 
The resulting Hamiltonian takes the form
\begin{equation}
	H(k)+\Delta_H^{\rm sub}(k)=
	\begin{pmatrix}
		\eta_H\cos k\,\sigma_x&Q(k)\\
		Q^\dagger(k)&0
	\end{pmatrix}.
\end{equation}
Clearly, since the upper diagonal block is no longer zero, the full SCS is broken. However, the lower diagonal block remains zero, and therefore
\begin{equation}
	(\mathcal S_H(H+\Delta_H^{\rm sub})\mathcal S_H^{-1}+H+\Delta_H^{\rm sub})P_{H,-}=0.
\end{equation}
In this way, the full SCS is reduced to a SubSy associated with the $\tau_-$ sector.

We next construct a second SubSy by selectively breaking ICS. To this end, we introduce a pseudospin-dependent hopping perturbation into the original non-Hermitian Hamiltonian $Q(k)$, given by $\eta_Q\cos k\,\sigma_-$. The modified non-Hermitian Hamiltonian is then written as
\begin{equation}
	Q(k)+\eta_Q\cos k\,\sigma_-=
	\begin{pmatrix}
		0&q_+(k)\\
		q_-^*(k)&\eta_Q\cos k
	\end{pmatrix}.
\end{equation}
Substituting the new non-Hermitian Hamiltonian into the doubling construction, the corresponding perturbation is
\begin{equation}
	\Delta_Q^{\rm sub}(k)=\eta_Q\cos k\,\tau_x\sigma_-.
\end{equation}
Unlike \(\Delta_H^{\rm sub}\), which modifies a diagonal block of the doubled Hamiltonian, \(\Delta_Q^{\rm sub}\) acts within its off-diagonal blocks.
Thus, $\Delta_Q^{\rm sub}$ alone preserves SCS but breaks the full ICS. 
Nevertheless, because the perturbation only involves the $B$ pseudospin subspace, it satisfies $\Delta_Q^{\rm sub}P_{Q,+}=0$, leading to
\begin{equation}
	(\mathcal S_Q(H+\Delta_Q^{\rm sub})\mathcal S_Q^{-1}+H+\Delta_Q^{\rm sub})P_{Q,+}=0.
\end{equation}
Hence, the full ICS is reduced to another SubSy, now associated with the $\sigma_+$ pseudospin subspace. The two SubSy conditions are therefore realized through distinct perturbations acting on different internal degrees of freedom.

Now we add both perturbations simultaneously to the original doubled Hamiltonian:
\begin{equation}
	H^{\rm sub}(k)=H(k)+\Delta_H^{\rm sub}(k)+\Delta_Q^{\rm sub}(k),
\end{equation}
and expanding it in the basis $(\tau_+\sigma_+,\tau_+\sigma_-,\tau_-\sigma_+,\tau_-\sigma_-)$, we obtain
\begin{equation}
	H^{\rm sub}(k)=
	\begin{pmatrix}
		0&\eta_H\cos k&0&q_+(k)\\
		\eta_H\cos k&0&q_-^*(k)&\eta_Q\cos k\\
		0&q_-(k)&0&0\\
		q_+^*(k)&\eta_Q\cos k&0&0
	\end{pmatrix}.
\end{equation}
When both $\eta_H$ and $\eta_Q$ are nonzero, neither the full SCS nor the full ICS is preserved. Nevertheless, the Hamiltonian simultaneously satisfies the two projected symmetry constraints,
\begin{equation}
	\begin{aligned}
			(\mathcal S_HH^{\rm sub}\mathcal S_H^{-1}+H^{\rm sub})P_{H,-}=0,\\
			(\mathcal S_QH^{\rm sub}\mathcal S_Q^{-1}+H^{\rm sub})P_{Q,+}=0.
	\end{aligned}	
	\label{eq:subsys}
\end{equation}
These constraints hold throughout the parameter range considered below, without requiring any modification of the perturbation strengths or the projection operators.

\section{Edge-state Switching under Subsymmetry-Preserving Disorder}
To further investigate the robustness of the symmetry-protected edge states and their switching between different boundaries, we introduce spatially random on-site perturbations that preserve both subsymmetries.
Specifically, we replace the translationally invariant hopping perturbations in Eq. (13) in the main text by random on-site couplings, while retaining their original internal matrix structure. 
In the real-space basis $(\tau_+ \sigma_+,\tau_+ \sigma_-,\tau_- \sigma_+,\tau_- \sigma_-)$, the resulting OBC Hamiltonian takes the form
\begin{equation}
	H_{\mathrm{OBC}}(V,W)=
	\begin{pmatrix}
		0&D_H&0&F\\
		D_H&0&G&D_Q\\
		0&G^{\dagger}&0&0\\
		F^\dagger&D_Q&0&0
	\end{pmatrix},
\end{equation}
where $G$ and $F$ are defined as in Eq.~\eqref{Seq:FG}.
The disorder matrices are defined as
\begin{equation}
	D_\alpha=W\,\mathrm{diag}(r_{\alpha,1},r_{\alpha,2},\ldots,r_{\alpha,L}),
	\qquad \alpha=H,Q,
\end{equation}
where $r_{\alpha,j}$ are independent random variables uniformly distributed in $[-1,1]$. Thus, $W$ controls the disorder strength, and the local coupling amplitudes independently vary within $[-W,W]$.
Moreover, the disorder terms satisfy
\begin{equation}
	P_{H,-}H_{\mathrm{dis}}P_{H,-}=0,\qquad
	P_{Q,+}H_{\mathrm{dis}}P_{Q,+}=0,
\end{equation}
and hence preserve the two corresponding subsymmetry constraints, even though translational invariance is broken.

Fig.~\ref{fig:figures2} illustrates the resulting OBC spectra.
In Fig.~\ref{fig:figures2}(a), the low-energy spectrum is shown as a function of $V$ for a fixed disorder strength $W=0.4$.
The edge-localized states remain clearly distinguishable in the different topological regimes.
Figs.~\ref{fig:figures2}(b) to ~\ref{fig:figures2}(d) further show the evolution of the spectra with increasing disorder strength for three representative parameter values.
Zero-energy edge modes remain in each of the three regimes within the displayed disorder range.
These results demonstrate the robustness of the symmetry-selected edge states and their switching between edges under SubSy-preserving disorder.

\begin{figure}
	\centering
	\includegraphics[width=0.8\linewidth]{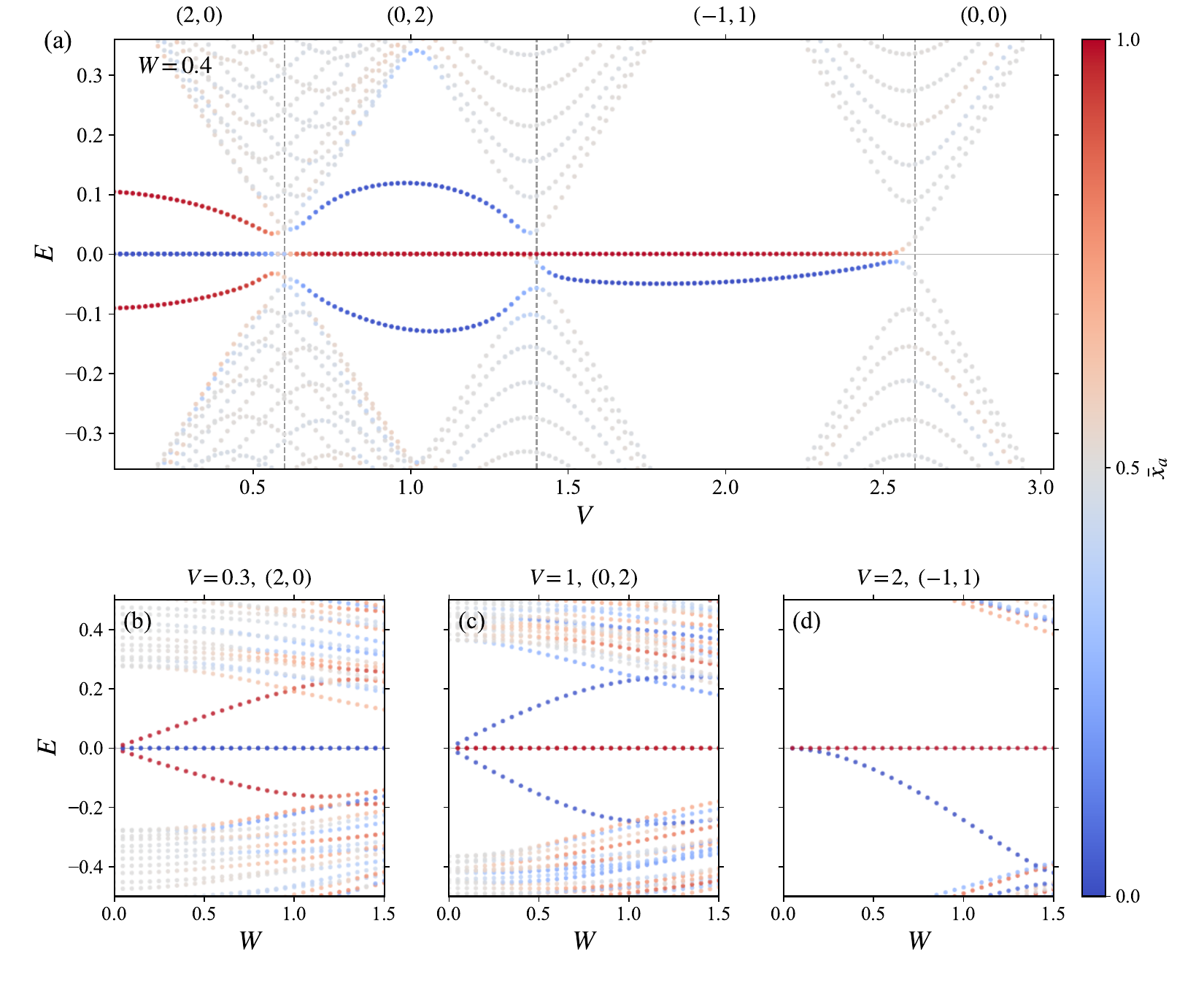}
	\caption{(a) Low-energy OBC spectrum as a function of $V$ at a fixed disorder strength $W=0.4$. The color represents the normalized mean position of each eigenstate. Blue and red indicate states localized near the left and right boundaries, respectively.
	 The phase labels $(\nu_H,\nu_Q)$ indicate the topological invariants of the corresponding clean system. 
	 The vertical dashed lines mark the clean-limit bulk gap-closing points at $V=0.6$, $1.4$, and $2.6$. 
	 (b)--(d) Low-energy OBC spectra as functions of the disorder strength $W$ for $V=0.3$, $1$, and $2$, respectively. 
	 The same disorder realization is used throughout, with its overall amplitude scaled by $W$. The system contains $L=200$ unit cells.}
	\label{fig:figures2}
\end{figure}

\end{widetext}

\bibliography{references}

\end{document}